\documentclass[notitlepage]{revtex4-1}

\usepackage{amssymb,amsfonts,amsmath}
\usepackage{graphicx}
\usepackage{epstopdf}
\usepackage{color}
\usepackage{eurosym}

\begin{document}

\title{Credit card fraud detection through parenclitic network analysis}
\author{Massimiliano Zanin$^{1}$, Miguel Romance$^{2,3}$, Santiago Moral$^{2,4}$  and Regino Criado$^{2,3}$}
\affiliation{$^{1}$ Department of Computer Science, Faculty of Science and Technology, Universidade Nova de Lisboa, Lisboa, Portugal}
\affiliation{$^{2}$ Department of Applied Mathematics, Universidad Rey Juan Carlos,  28933 M\'ostoles, Madrid, Spain}
\affiliation{$^{3}$ Center for Biomedical Technology, Universidad Polit\'ecnica de Madrid,  28223 Pozuelo de Alarc\'on, Madrid, Spain}
\affiliation{$^{4}$ Cyber Security \& Digital Trust, BBVA Group,  28050 Madrid, Spain}

\date{\today}

\begin{abstract}
The detection of frauds in credit card transactions is a major topic in financial research, of profound economic implications. While this has hitherto been tackled through data analysis techniques, the resemblances between this and other problems, like the design of recommendation systems and of diagnostic / prognostic medical tools, suggest that a complex network approach may yield important benefits.
In this contribution we present a first hybrid data mining / complex network classification algorithm, able to detect illegal instances in a real card transaction data set. It is based on a recently proposed network reconstruction algorithm that allows creating representations of the deviation of one instance from a reference group.
We show how the inclusion of features extracted from the network data representation improves the score obtained by a standard, neural network-based classification algorithm; and additionally how this combined approach can outperform a commercial fraud detection system in specific operation niches.
Beyond these specific results, this contribution represents a new example on how complex networks and data mining can be integrated as complementary tools, with the former providing a view to data beyond the capabilities of the latter.
\end{abstract}

\maketitle

\section{Introduction}\label{sec:intro}

Credit card frauds, a concept included in the wider notion of financial frauds \cite{ngai2011application, west2016intelligent}, is a topic attracting an increasing attention from the scientific community. This is due, on the one hand, to the raising costs that they generate for the system, reaching billions of dollars in yearly losses and a percentage loss of revenues equal to the $1.4\%$ of online payments \cite{bhatla2003understanding}. On the other hand, credit card frauds have important social consequences and ramifications, as they support organised crime, terrorism funding and international narcotics trafficking - see \cite{rollins2006terrorist} for a complete review.

Detecting unauthorised credit card transactions is an extremely complex problem, as features are seldom useful if taken individually. To illustrate, a large transaction is not {\it prima facie} suspicious, unless it is performed at usual times ({\it e.g.} at night) or in an unusual store ({\it i.e.} a store never visited before by the card owner, located in a different city, {\it etc.}).
When different features have to be combined in non-trivial ways, the customary solution is to resort to data mining, a sub-field of computer science dealing with the automatic discovery of patterns in data sets \cite{friedman2001elements, han2011data, witten2016data}.
Most of the data mining models to detect credit card frauds are based on artificial neural networks (ANN), a model inspired on the structural aspects of biological neural networks, and in which a set of nodes process the input signal by interacting between them \cite{zurada1992introduction, hagan1996neural}. This does not come as a surprise, as ANN are able to extract complex non-linear patterns from data, with almost no hypotheses on the underlying structure. ANN yielded good results in credit card classification tasks, as for instance in \cite{ghosh1994credit, aleskerov1997cardwatch, brause1999neural, maes2002credit, syeda2002parallel, carneiro2015cluster} - see \cite{bhattacharyya2011data, sethi2014revived, zojaji2016survey} for reviews.

While data mining algorithms are able to detect hidden patterns in data, they usually lack the capacity of synthesising metrics describing the global structure created by the interactions between the different features. In recent years, the use of complex networks theory has been proposed as a way of overcoming this limitation. Complex networks are a statistical-mechanics understanding of the classical graph theory, aimed at describing and characterising the structure of complex systems \cite{strogatz2001exploring, albert2002statistical, boccaletti2006complex}. The interaction between network theory and data mining is bidirectional: the former can be used to synthesise high-level features to be fed into a classification problem; while the latter can endow networks with an objective way of validating results - see \cite{zanin2016combining} for a complete review.

More specifically, complex networks and data mining can be integrated as complementary tools in order to extract, synthesise and create new representations of a data source, with the aim of, for instance, discover new hidden patterns in a complex structure. The appropriate integration of complex network metrics can result in improved classification rates with respect to classical data mining algorithms and, reciprocally, there are many situations in which data mining can be used to solve important issues in complex network theory and applications \cite{zanin2016combining}.

In this contribution we explore the possibility of using complex networks as a way of improving credit card fraud detection. Specifically, networks are used to synthesise complex features representing card transactions, relying on the recently proposed approach of {\it parenclitic networks} (Section \ref{sec:methods}). Afterwards, their relevance is evaluated by means of a large dataset of real transactions, by comparing the yielded increase in the classification score when compared to the use of a standard ANN algorithm (Section \ref{sec:results}). We additionally show that the combined data mining / complex networks approach is able to outperform a commercial system in some specific situations.

\section{Methods}\label{sec:methods}

In this section we present the main tools that are going to be used for the classification of credit card transactions between licit and illicit. Given a credit card transaction $t_i$ with features $f_{i1},\cdots,f_{ik}$, the problem entails detecting if it is illicit or not from its features and the knowledge obtained from an historical training dataset - what is known as a supervised learning problem. From a mathematical point of view, we have to model a function $H:\mathbb{R}^k\longrightarrow\mathbb{R}$ and find $\delta>0$ such that if $|H(f_{i1},\cdots,f_{ik})|\le \delta$, then $t_i$ is not illicit. Note that, while there are multiple types of illicit patterns, such aspect is here not considered, in that any suspicious transaction is considered as a potential fraudulent one.


We firstly introduce the concept of {\it parenclitic networks} in Section \ref{sec:parenclitic}, a network reconstruction technique that allows highlighting the differences between one instance and a set of standard ({\it i.e.} baseline, or in this case licit) instances \cite{zanin2011complex, zanin2014parenclitic}. We subsequently describe the real data set used for validation (Section \ref{sec:dataset}), including the available raw features (Table \ref{tab:Features}); and the global classification model (Section \ref{sec:ClassModels}).

\subsection{Parenclitic networks reconstruction}\label{sec:parenclitic}


As initially proposed in \cite{zanin2011complex}, one may hypothesise that the right classification of an observation does not only come from its features, but also from the structure of correlations between them. Following the mathematica formalism introduced before, if we consider the set 
\[
L=\left\{(x_1,\cdots,x_k)\in\mathbb{R}^k;\enspace |H(x_1,\cdots,x_k)|\le \delta\right\}\subseteq\mathbb{R}^k,
\]
then $L$ is a manifold in $\mathbb{R}^k$ such that if we take a (new) transaction $t$ with features $t_1,\cdots,t_k$ such that $(t_1,\cdots,t_k)\notin L$, then $t$ is considered as an illicit transaction. In general it is computationally impossible to obtain the set $L$ directly from the training dataset, since it is a high dimensional problem. As an alternative, the parenclitic approach analyses the family of projections of $L$ into 2-dimensional spaces corresponding to couples of features $(x_i,x_j)$ with $1\le i\ne j\le k$. Hence, if we consider a training dataset with $n\in\mathbb{N}$ transactions, each of them described by $k\in\mathbb{N}$ (numeric) features, we can analyse up to ${k\choose 2}=k(k-1)/2$ two-dimensional projections of pairs of different features, each of them with up to $n$ points in $\mathbb{R}^2$. In order to quantify the correlation between pairs of features, the parenclitic approach proposes associating a network to each transaction with $k$ nodes (as many as features considered) and the links measure the correlation between features \cite{zanin2014parenclitic}. Hence the following pre-processing must be completed: for every two-dimensional projection of $L$ given by a couple of features $(f_i,f_j)$ with $1\le i\ne j\le k$, the correlation for the licit transactions in the training dataset is measured (by means of, for instance, a linear regression or other curve fitting techniques). For the shake of simplicity, we have here considered a linear regression, such that every pair of features $(f_i,f_j)$ with $1\le i\ne j\le k$ yields a linear fitting between $f_i$ and $f_j$ for the licit transactions in the training dataset. Mathematically, this is represented by a linear equation of the form:
\[
r_{ij}:\,x_j=a_{ij}x_i+b_{ij}.
\]

Once these ${k\choose 2}$ linear regression lines are computed, a threshold $\alpha>0$ is fixed. Given a new ({\it i.e.} not included in the training set) transaction $t$ with features $t_1,\cdots,t_k$ , a network $G=G(t)$ is associated to $t$ as follows:

\begin{itemize}
 \item $G$ has $k$ nodes $1,\cdots,k$,
 \item For every pair of nodes $1\le i\ne j\le k$ we compute $w_ij\ge 0$  as the (euclidian) distance from $(t_i,t_j)$ to the line $r_{ij}$ in $\mathbb{R}^2$, i.e.
 \[
 w_{ij}=d((t_i,t_j),r_{ij}).
 \]
 As an alternative, the euclidian distance could be replaced by any pseudo-distance function in $\mathbb{R}^2$. For the shake of simplicity, the euclidian distance will be used in this paper, but similar results can be obtained for other pseudo-distance functions.
 \item For every pair of nodes $1\le i\ne j\le k$, the (undirected) link $(i,j)$ is in graph $G$ if and only if $w_{ij}\ge \alpha$.
\end{itemize}

Note that the parenclitic network $G(t)$ summarises the couples of features whose correlation strongly differs from a typical licit transaction; the structure of this network thus contains valuable information about the (abnormal) correlation of features in the credit card transaction. Once this parenclitic network is computed, it is necessary to transform it in a set of features compatible with a data mining algorithm. Towards this end, several structural measures have been extracted, and will be considered as new features associated with the transaction (see next section for details). Among all possible structural measures that could be computed (see, for example, Ref.~\cite{costa2007characterization} and references therein), those here selected are summarised in Table~\ref{table:measures}.

\begin{table}[!tb]
\begin{center}
\begin{tabular}{ | p{4cm} | p{10cm} | }
  \hline 
  Name & Description \\ 
  \hline
  Maximum node degree~\cite{costa2007characterization} & Maximum degree of all nodes in the network. It is calculated as $M_k = \max_i k_i$, $k_i$ being the degree of nodes $i$ \\
  Entropy of the degree distribution~\cite{wang2006entropy} & Shannon entropy of the distribution of nodes degrees. It is given by $E = - \sum _{i=0} ^{M_k} p_i \log p_i$, $p_i$ being the probability of finding a node of degree $i$. \\
  Assortativity~\cite{costa2007characterization} & Pearson's correlation coefficient between the degree of connected nodes. \\
  Clustering coefficient~\cite{costa2007characterization} & Measure of the presence of triangles in the network. It is defined as the number of triangles (groups of three fully-connected nodes) over the number of connected triplets (groups of three nodes connected by at least two links). \\
  Geodesic distance~\cite{costa2007characterization} & Average length of the shortest path connecting pairs of nodes. \\
  Efficiency~\cite{latora2001efficient} & Inverse of the harmonic mean of the length of all shortest distances. \\
  Information Content~\cite{zanin2014information} & Metric assessing the presence of meso-scale structures in the network. \\ 
  \hline
  \end{tabular}
\end{center}
\caption{\label{tab:Topological} List of topological metrics used to describe the structure of parenclitic networks.}\label{table:measures}
\end{table}

\subsection{Data set description}\label{sec:dataset}

The data set here considered includes all credit and debit card transactions of clients of the Spanish bank BBVA, from January 2011 to December 2012. Each month, an average of $15$ million operations were realized by $7$ million cards, for a total of $250$ GB of information.  

Transactions are automatically screened by an algorithm designed to detect suspected transactions, and returning a score from $0$ (no suspect) to $100$ (potentially illegal). Afterwards, transactions are classified in two categories, {\it i.e.} {\it legal} and {\it illegal}, as the result of a manual classification performed by the bank's legal personnel - using both information of the automatic algorithm, and customers' complaints. This allows us to detect which transactions were positively detected as frauds by the automatic algorithm, and which were false negatives.

Available fields included a time stamp of the operation, the quantity (both in Euro and in the original currency, if different), and the origin (the card) and destination (the store) of the operation; the two latter fields were anonymised, so that the exact card number and the name of the store could not be recovered. Some additional features have been synthesised from the previous ones, {\it e.g.} the average transaction size of a given user. A full list of the available fields is reported in Tab~\ref{tab:Features}. Additionally, a full statistical characterisation of the features can be found in Ref.~\cite{Zanin2016}, including the temporal evolution of the structure of the transactions network.

\begin{table}[!tb]
\begin{center}
\begin{tabular}{ | l | l | p{10cm} | }
  \hline 
  Name & Type & Description \\ \hline
  Transaction size & Integer & Size, in Euro, of the transaction under analysis. \\
  Time since last transaction & Integer & Time, in seconds, since the last transaction of the same card. \\
  Last transaction size & Integer & Size, in Euro, of the previous transaction executed by the same card. \\
  Average transaction size & Float & Average size, in Euro, of the transactions executed by the card in the last month. \\
  Average time between transactions & Float & Average time, in seconds, between consecutive transactions of the same card. \\
  Same shop & Boolean & $1$ is the shop corresponds to the one of the last transaction of the same card, $0$ otherwise. \\
  Hour of the day & Integer & Hour (from $1$ to $24$) at which the operation was realised. \\
  Fraud rate & Float & Average rate of illegal operations, for all cards, in the last $50.000$ transactions. \\ \hline
  Fraud suspectness & Integer & Number representing the likelihood for the transaction to be illicit, according to the bank automatic fraud detection algorithm. Values range between $0$ (no fraud suspected) to $100$ (certain fraud). \\
  Fraud & Boolean & $1$ if the transaction has been recognised as a fraud, $0$ otherwise. \\ \hline
\end{tabular}
\end{center}
\caption{\label{tab:Features}Features composing the credit card transactions dataset.}
\end{table}

\subsection{Classification models}\label{sec:ClassModels}

As previously introduced, in this contribution we are going to explore two different ways of detecting illicit credit card transactions: a classical data mining approach, and the introduction of features extracted from a network representation. In both cases, the process must follow some common steps: it is first necessary to extract the expected behavior, {\it i.e.} a set of features representing the typical legal and illegal transaction; for then building a model that learns from those features, and yields an expected classification for a new transaction not yet studied.

\begin{figure}[!tb]
\begin{center}
\includegraphics[width=0.5\textwidth]{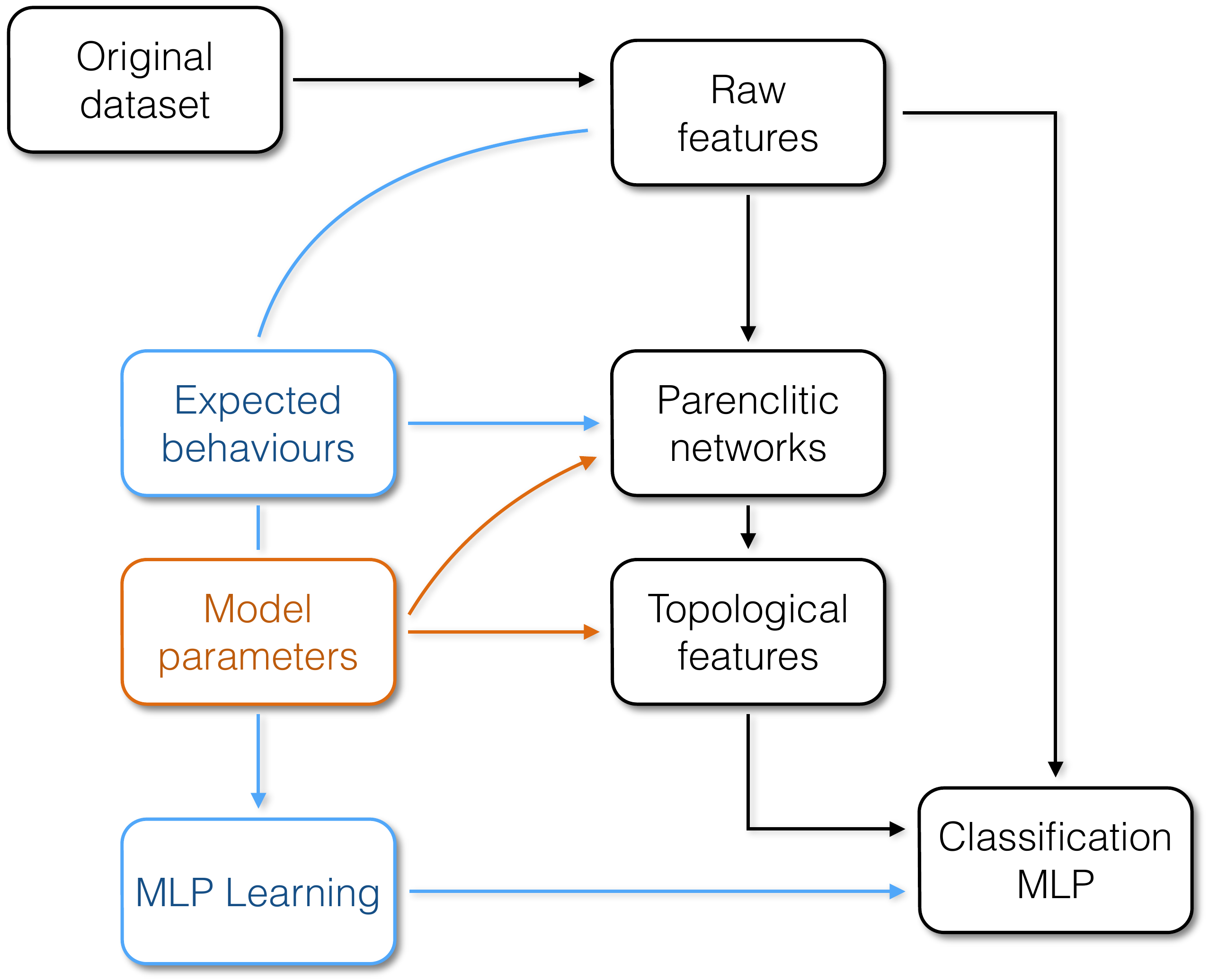}
\end{center}
\caption{\label{fig01} Schematic representation of the classification model. See main text for details.}
\end{figure}

Fig.~\ref{fig01} depicts an overview of the whole process. It starts from the original data set, from which a set of raw features are extracted - as described in Section \ref{sec:dataset} and listed in Tab. \ref{tab:Features}. The features corresponding to the licit transactions are then used to recover the normal relations, as described in Section \ref{sec:parenclitic}, and to reconstruct the parenclitic networks of all transactions. These networks are then binarised, {\it i.e.} links with weight below a given threshold are deleted, and a set of topological metrics are extracted - see Table~\ref{tab:Topological} for a complete list. Note that, at the end of this analysis, all transactions are described by $15$ features: $8$ coming from the raw data, and $7$ from the network analysis.

Artificial Neural Networks (ANNs), and specifically Muti-Layer Perceptrons (MLP) have been chosen as the final model for classifying new transactions. They are inspired by the structural aspects of biological neural networks, and are represented by a set of connected nodes in which each connection has a weight associated with it, and the network learns the classification function adjusting the node weights~\cite{rosenblatt1958perceptron, hagan1996neural}. The output of each artificial neuron $j$ is defined by: 

\begin{equation}
	f(W^T, x_j)=\sum^{n}_{i=1}{W_i x_i + b},
\end{equation}

$W$ being the vector of weights, and $f$ the sigmoid activation function:

\begin{equation}
f(x)= \frac{1}{1 + \exp(-x)}.
\end{equation}

Following the standard configuration, neurons were organized in three layers: an input one, with a number of neurons equal to the input features; an intermediate, or hidden one, with ten neurons; and a final output layer comprising just one computational element. The training has been performed with the standard back-propagation algorithm~\cite{Werbos74}. Finally, the reconstruction of the MLP models has been performed using the KNIME software \cite{berthold2009knime}.

The evaluation of the classification efficiency has been performed using both sensitivity (also known as True Positive Rate - TPR) and Receiver Operating Characteristic (ROC) curves \cite{hanley1982meaning}. These curves are created by plotting the True Positive Rate (TPR) against the False Positive Rate (FPR) at various threshold settings. ROC plots present the important advantage of showing the performance of the classification model for different sensitivity values. This is relevant for the problem at hand, as false positives are extremely expensive, {\it e.g.} in terms of the negative commercial image of the bank; conservative solutions are therefore usually preferred.

\section{Results}
\label{sec:results}

As explained in Section \ref{sec:parenclitic}, the parenclitic approach usually requires the definition of a threshold $\alpha$, which is used to binarise the (initially weighted) networks. Instead of using an {\it a priori} approach, {\it i.e.} the definition of $\alpha$ using expert judgement, we here tackle the problem indirectly, by following the procedure proposed in Ref. \cite{zanin2012optimizing}. Specifically, we optimise the network reconstruction by finding the link density (and hence the value of $\alpha$) that optimises the efficacy of the classification model.

\begin{figure}[!tb]
\begin{center}
\includegraphics[width=0.95\textwidth]{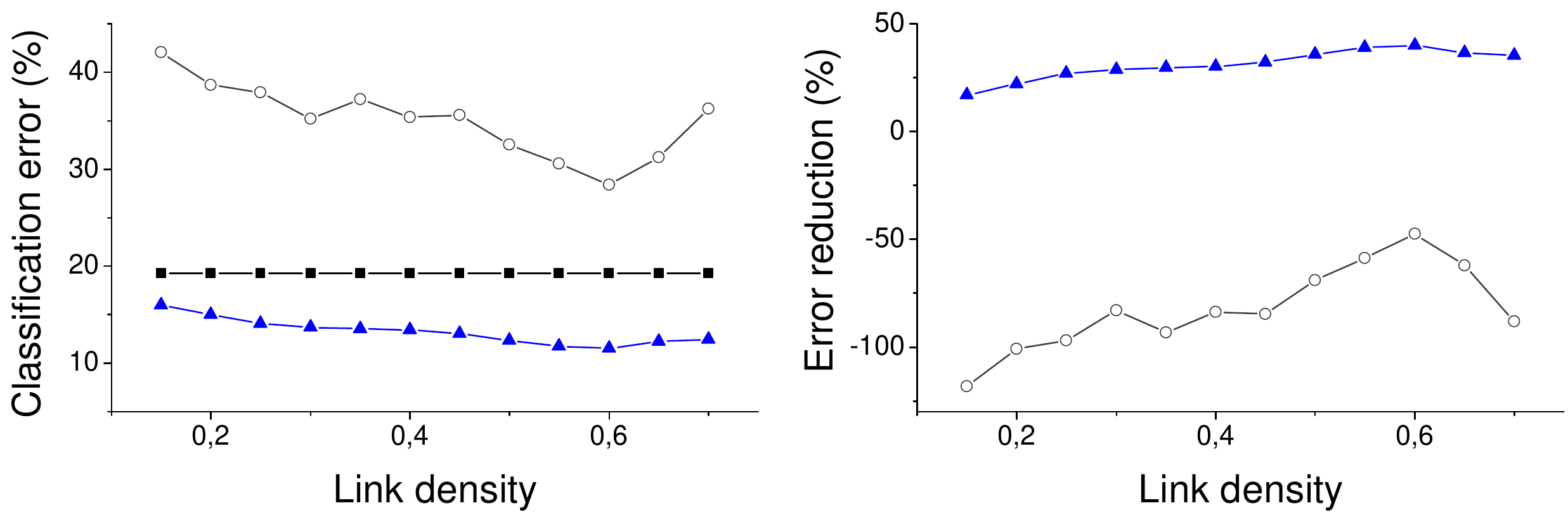}
\end{center}
\caption{\label{fig02} (Left) Classification error as a function of the link density of the parenclitic networks. Black squares, black circles and blue triangles respectively represent the error for the classification for the original raw features, for the classification using parenclitic features alone, and for the classification with all features. (Right) Error reduction, in percentage, when using only parenclitic features (black circles) and the full set of features (blue triangles), w.r.t. the use of the raw data set.}
\end{figure}

Fig.~\ref{fig02} Left presents the evolution of the classification error (sensitivity or TPR) as a function of the considered link density, for three different scenarios: the use of only the raw features, as described in Tab.~\ref{tab:Features} (solid black squares); the use of the features extracted from the parenclitic representation alone (hollow black circles); and the use of the combined sets of features (solid blue triangles). Note that, in the former case, the result is constant, as the original features are not affected by the binarisation process. In order to avoid overfitting, this classification has been performed on a balanced sub data set, composed of an equal number of legal and illegal transactions.

Several conclusions can be drawn from Fig.~\ref{fig02}. First of all, the features extracted from the parenclitic networks are not enough, alone, to reach a low classification error. This has to be expected: while important information can be codified in the interaction between raw features, some important clues may be hidden in the latter, {\it e.g.} abnormal transaction sizes or timings. At the same time, the addition of parenclitic features to the raw data set enhance the obtained results, with the error dropping from a $19.2\%$ to a $12.23\%$. This is further illustrated in Fig.~\ref{fig02} Right, depicting the reduction in the classification error (in percentage) when considering only parenclitic features and the whole data set - note that, in the first case, the reduction is negative as the error increases. Finally, the best classification suggests that the optimal link density that should be considered is of a $60\%$ - meaning that the $40\%$ of links with less weight should be deleted.

\begin{figure}[!tb]
\begin{center}
\includegraphics[width=0.4\textwidth]{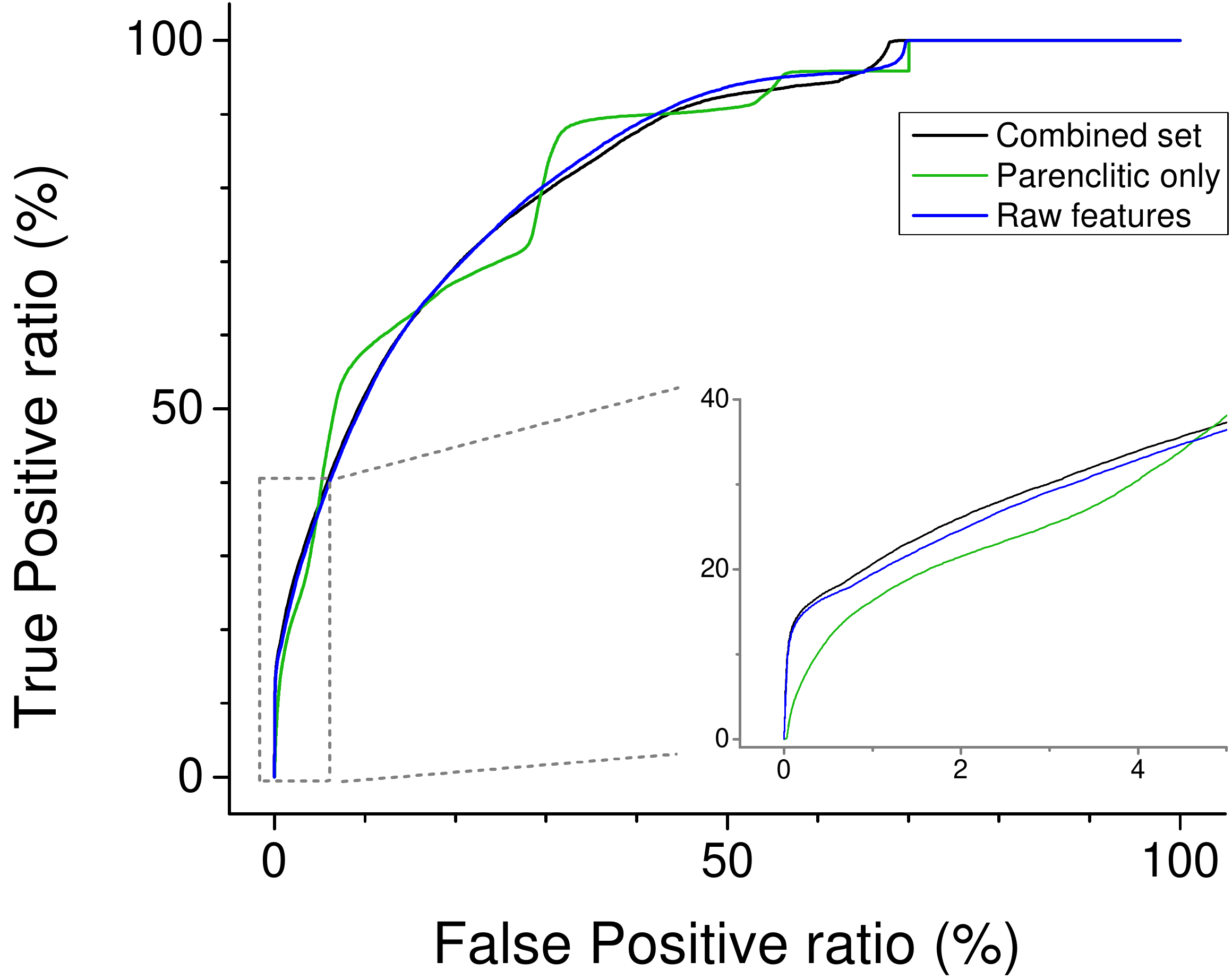}
\hspace{0.5cm}
\includegraphics[width=0.4\textwidth]{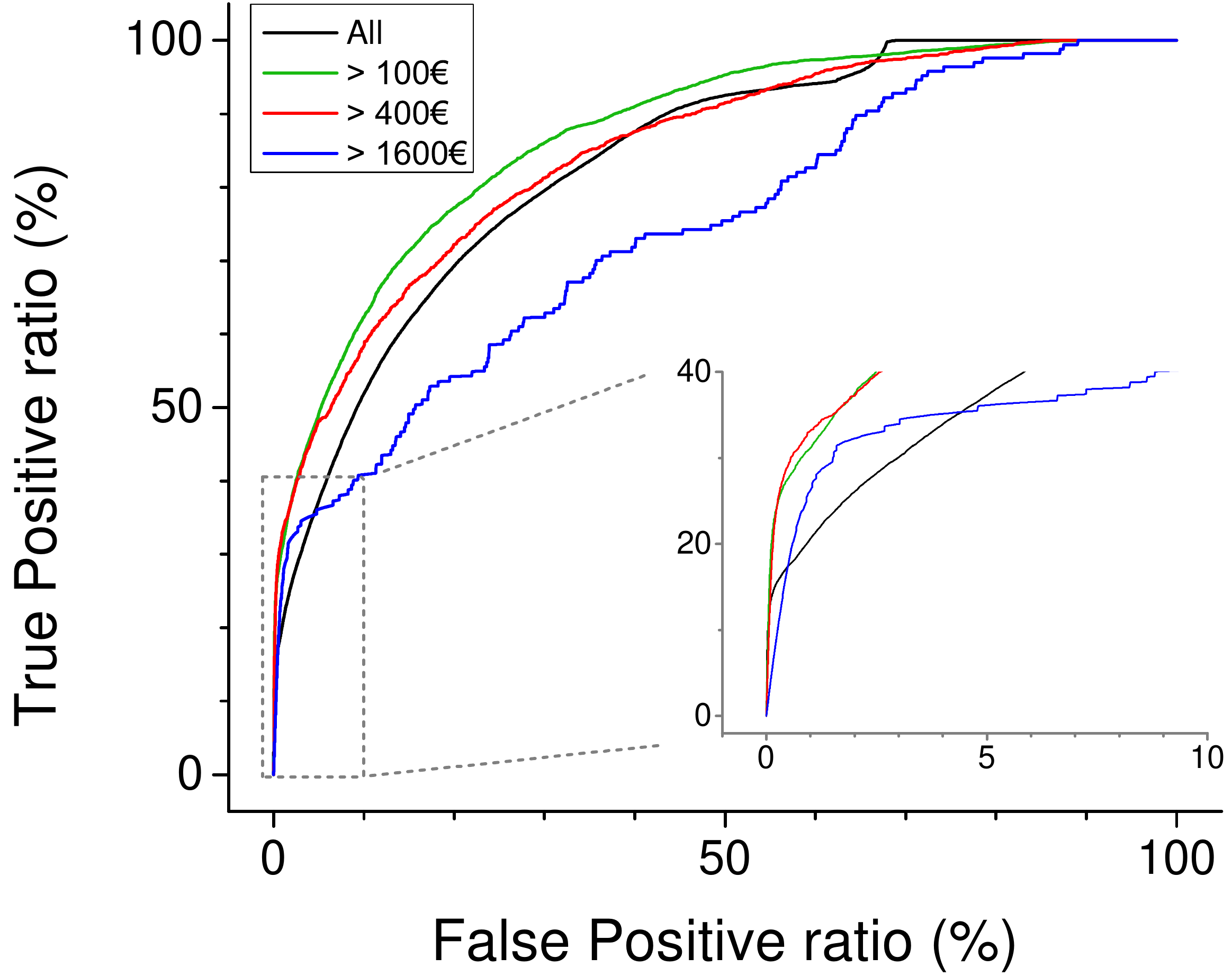}
\end{center}
\caption{\label{fig03} (Left) ROC curves of the classification, corresponding to the use of the raw features alone (blue line), of the parenclitic features (green line), and of the combined sets (black line). (Right) ROC curves, obtained through the combined data set, as a function of the transaction sizes.}
\end{figure}

If Fig.~\ref{fig02} is useful to detect the best link density for the analysis, it does not convey information about the real performance of the classification algorithm in an operational environment. For that, Fig.~\ref{fig03} Left presents three ROC curves, corresponding to the use of raw (blue line), parenclitic (green line), and combined features (black line) as before. Note that results here presented correspond to the optimal link density of $60\%$, as previously estimated.
As previously discussed, the most interesting operational configuration is the one minimising the number of false positives, as this minimises the commercial costs of the organisation. The inset of Fig.~\ref{fig03} thus shows the bottom left part of the curves. It  can be appreciated that, after an initial part in which results are comparable, the addition of the parenclitic features slightly increases the number of true positives - note how the black line is above the blue one.
Even though this may seem a negligible difference, it is worth noting that any improvement, however small, has a significant impact due to the large number of transactions managed by the system. Increasing the fraud detection rate by $1\%$ would allow identify $\approx 20.000$ new illicit transactions per year, or $\approx 2$M\euro~in saved costs.

\begin{figure}[!tb]
\begin{center}
\includegraphics[width=0.40\textwidth]{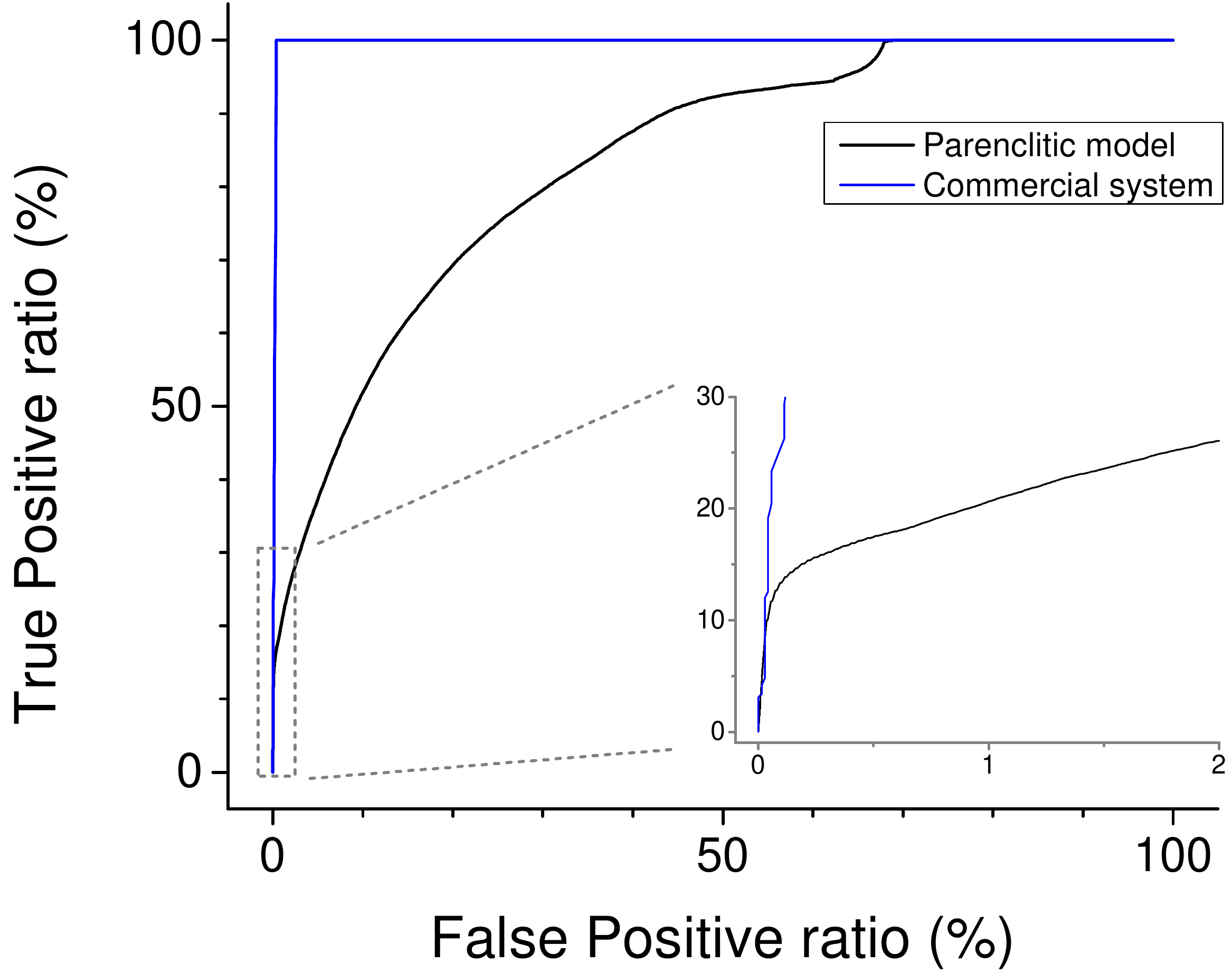}
\hspace{0.5cm}
\includegraphics[width=0.40\textwidth]{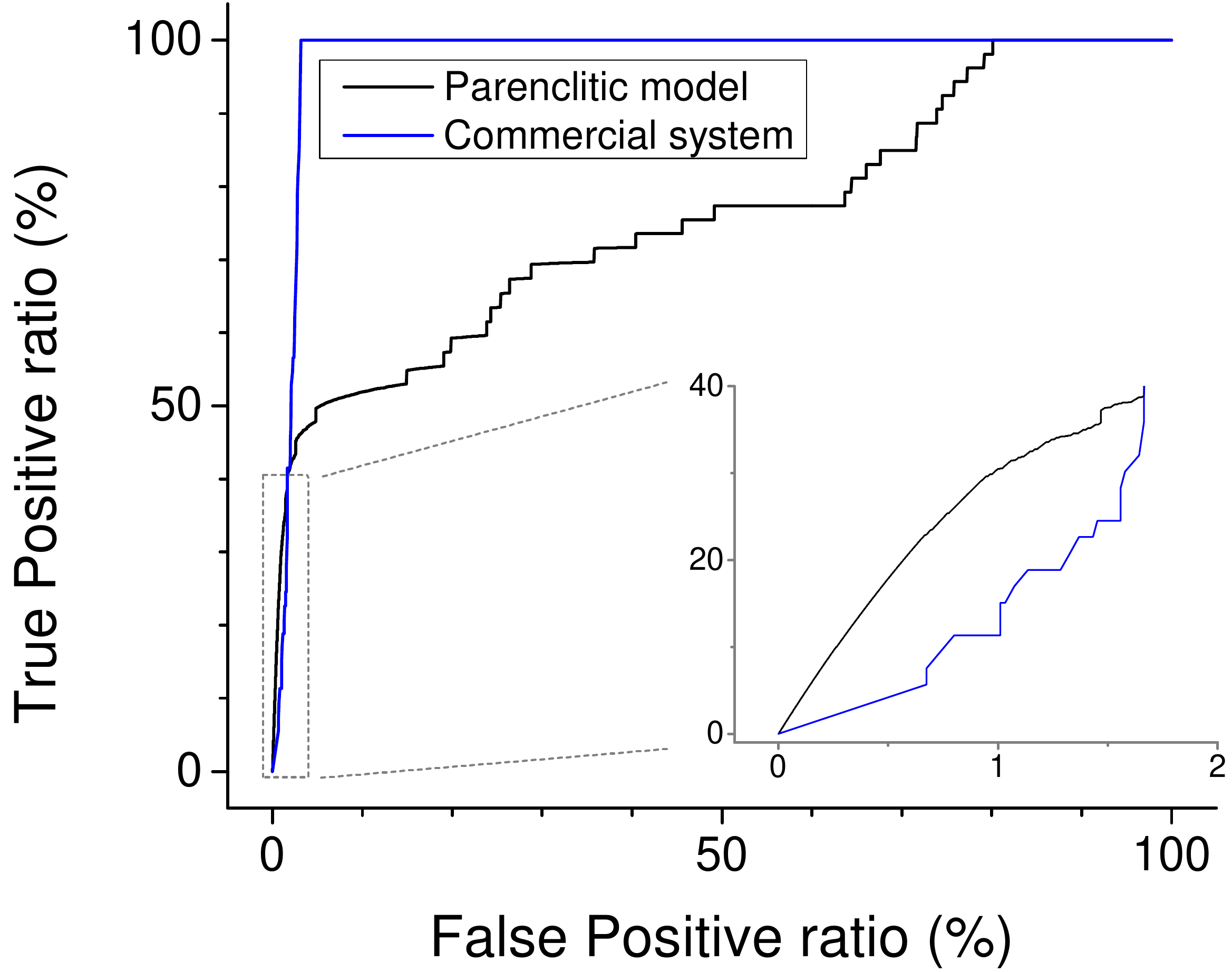}
\end{center}
\caption{\label{fig04} (Left) ROC curves for the network-based model (black line) and a commercial system (blue line). (Right) ROC curves, for the proposed network-based algorithm and a commercial system, when only on-line (Internet) transactions are considered.}
\end{figure}

Fig. \ref{fig03} Right further presents four ROC curves calculated for different transaction sizes: all transactions (black line), and transactions above $100$\euro~(green line), $400$\euro~(red line) and $1.600$\euro~(blue line). Deleting small transactions results in an improvement of the detection efficiency - note how the green and red lines lay above the black one. Additionally, the proposed algorithm fails for large transactions; this does not come as a surprise, as the larger the size, the fewer the available instances, making training more challenging.

If what previously presented illustrates that the use of a network representation can improve a fraud detection algorithm, it does not clarify how it ranks against a commercial system. As may be expected, the proposed algorithm is less efficient than the fraud score included in the original data set - see Fig. \ref{fig04} Left\footnote{Due to confidentiality issues, the name and characteristics of the commercial fraud detection system cannot be included in this publication.}. Nevertheless, there are niches in which the opposite happens, the most important being the analysis of on-line transactions. Fig. \ref{fig04} Right depicts two ROC curves, respectively for the algorithm based on parenclitic networks (black line) and the commercial system (blue line), when only transactions realised through Internet are considered. While the commercial system clearly outperforms the proposed algorithm, with an Area Under the Curve (AUC) close to $1.0$, the latter is slightly better for a low ratio of False Positive - as previously explained, the plane region most interesting for real operations.

\section{Conclusions}

Complex networks and data mining models share more characteristics that what may {\it prima facie} appear, most notably having similar objectives: both aim at extracting information from (potentially complex) systems to ultimately generate new compact quantifiable representations. At the same time, they approach this common problem from two different approaches: the former by extracting and quantitatively evaluating the underlying structure, the latter by creating predictive models based on historical data \cite{zanin2016combining}.
In this contribution we test the hypothesis that complex networks can be used as a way to improve data mining models, framed within the problem of detecting fraud instances in credit card transactions, providing a new example about how complex networks and data mining may be integrated as complementary tools in a synergistic manner in order to improve  the classification rates obtained by classical data mining algorithms.

Results confirm that features extracted from a network-based representation of data, leveraging on a recently proposed parenclitic approach \cite{zanin2011complex, zanin2014parenclitic}, can play an important role: while not effective in themselves, such features can improve the score obtained by a standard ANN classification model.
We further show how the resulting model is especially efficient in detecting frauds in some niches of operations, like medium-sized and on-line transactions. Finally, we illustrate as, in the latter case, the network-based model is able to yield better results than a commercial fraud detection system. All results have been obtained with a unique data set, comprising all transactions managed during two years by a major Spanish bank, and including more than $180$ million operations.

\bibliography{Fraud}

\begin{acknowledgments}
Authors gratefully acknowledge the Technological Risk Management Research Center (Centro para la Gesti\'on Tecnol\'ogica del Riesgo, CIGTR) sponsored by the Rey Juan Carlos University and BBVA group, and the I4S-URJC Chair on Information Security, Fraud Prevention and Technological Risk Management that encourage the base research on pain points of risk management within information systems. 
 
This work has been partly supported by the Spanish MINECO under project  MTM2014-59906-P and by the grant for the researching activity for excellence group GARECOM GI\_EXCELENCIA 30VCPIGI11.

\vspace{0.5cm}
{\bf Author Contributions}
M.Z. conceived and elaborated the method, and performed the numerical experiments.  M.Z., M.R., R.C. and S.M. analysed the data, prepared the figures, and wrote the text of the Manuscript. All Authors reviewed the Manuscript.

\vspace{0.5cm}
{\bf Competing Interests}
The Authors declare no competing financial interests.

\end{acknowledgments}

\end{document}